# The Impact of TaS$_2$-Augmented Interconnects on Circuit Performance: A Temperature-Dependent Analysis

Xinkang Chen, and Sumeet Kumar Gupta, Senior *Member,* IEEE

*Abstract*—Monolayer TaS$_2$ is being explored as a future liner/barrier to circumvent the scalability issues of the state-of-the-art interconnects. However, its large vertical resistivity poses some concerns and mandates a comprehensive circuit analysis to understand the benefits and trade-offs of this technology. In this work, we present a detailed temperature-dependent modeling framework of TaS$_2$-augmented copper (Cu) interconnects and provide insights into their circuit implications. We build temperature-dependent 3D models for Cu-TaS$_2$ interconnect resistance capturing surface scattering and grain boundary scattering and integrate them in an RTL-GDSII design flow based on ASAP7 7nm process design kit. Using this framework, we perform synthesis and place-and-route (PnR) for advanced encryption standard (AES) circuit at different process and temperature corners and benchmark the circuit performance of Cu-TaS$_2$ interconnects against state-of-the-art interconnects. Our results show that Cu-TaS$_2$ interconnects yield an enhancement in the effective clock frequency of the AES circuit by 1%-10.6%. Considering the worst-case process-temperature corner, we further establish that the vertical resistivity of TaS$_2$ must be below 22 kΩ-nm to obtain performance benefits over conventional interconnects.

*Index Terms*—Back-end of the line (BEOL), interconnect, barrier, liner, grain boundary scattering, surface scattering, via.

## I. INTRODUCTION

Technology scaling has been among the most important drivers for the advancement of electronic systems and has led to unprecedented improvements in their speed, energy efficiency and integration density [1]. However, deeply scaled technologies exhibit several challenges [2], [3] that need to be addressed to sustain the benefits of scaling. The issues related to deeply scaled transistors such as short channel effects, increased leakage, higher parasitic capacitances are well known [3]. Several promising transistor designs offer a mitigation of such problems, which if successful, will continue to improve the transistor speed and energy efficiency. At the same time, to translate the benefits of transistor innovations to system performance, interconnect scaling assumes critical importance. Several studies have shown that the interconnects could become the performance bottlenecks in advanced technology nodes due to several scaling challenges [2].

First, scaling of interconnect width (or pitch) and height (to achieve lower capacitance and higher integration density) leads to increase in the resistance per unit length of the metal lines ($R_{METAL}$). Lower interconnect width also requires vias with smaller cross-section which increases their resistance ($R_{VIA}$). This is not only because of the direct impact of the geometry on resistance, but also due to the sidewall scattering, which significantly increases the resistivity (not just resistance) of the metal lines and vias with the scaling of their cross-sectional area [4]. To further aggravate the situation, the active conduction area of the state-of-the-art copper (Cu)-based interconnects is usually lower than their footprint. This is because Cu needs to be enclosed within barrier layers (e.g. TaN) to reduce electromigration [5], [6] and liners (e.g. Ta) for proper adhesion [5], [6]. This further increases the resistivity and resistance of the metal lines and vias. The impact of barrier/liner is aggravated with scaling as their thickness is difficult to scale.

To address these issues, several technological alternatives have been explored. Materials other than Cu, e.g. cobalt, which may not need barriers/liners is one option [7]. Processing techniques beyond the standard damascene process, such as subtractive processes [8], which can potentially avert the barrier/liners layers also look promising. However, these techniques typically trade-off the benefits of standard interconnects such as the maturity of the damascene process and high intrinsic conductivity of Cu. The question, therefore, arises, whether the active area of Cu can be increased for a given pitch by finding alternate barrier/liner materials whose thickness can be scaled aggressively.

To that end, 2D materials have been explored to replace TaN/Ta based barriers/liners [9]. Due to their intrinsic 2D nature, these materials seem to be a natural fit for this application. However, they must satisfy certain properties for easy integration with Cu, such as adhesion and the ability to prevent electromigration. Amongst different 2D materials, TaS$_2$ looks promising in this regard [9]. Monolayer TaS$_2$ films as low as 0.7nm can potentially replace *both* Ta and TaN, thus increasing the active conduction area of Cu. However, their vertical conductivity is small, which significantly increases the resistance of the vias, as discussed in detail later. Thus, while TaS$_2$ based liner/barrier can be effective in reducing $R_{METAL}$, the concomitant increase in $R_{VIA}$ offsets their potential benefits. In order to understand the overall implications of this promising

This work was supported , in part, by NSF and by the NEW materials for LogIc, Memory and InTerconnectS (NEWLIMITS) Center funded by the Semiconductor Research Corporation (SRC)/National Institute of Standards and Technology (NIST). (Corresponding author: Xinkang Chen.)
Xinkang Chen and Sumeet Kumar Gupta are with the School of Electrical and Computer Engineering, Purdue University, West Lafayette, IN 47907 USA (e-mail: chen3030@purdue.edu; guptask@purdue.edu).

technology, there is a need for a circuit-level evaluation, that can properly account for the two opposing effects of replacing TaN/Ta with $TaS_2$.

To that end, our previous works [10], [11] have analyzed the circuit implications of Cu-$TaS_2$ interconnects at a nominal process-temperature corner showing the circuit-level performance boost using this technology (compared to standard interconnects). In this paper, we expand upon our work by evaluating the temperature-dependence of $TaS_2$-augmented Cu interconnects and their impact on the performance of an interconnect-dominated benchmark – Advanced Encryption Standard (AES). For this, we model the conventional (TaN/Ta-based) and $TaS_2$-augmented interconnects using 3D COMSOL Multiphysics Modeling Software accounting for surface scattering [12], grain boundary scattering [13] and their temperature dependence. We incorporate the obtained interconnect parameters in a synthesis and place-and-route (PnR) flow based on ASAP7 standard cell library and benchmark the performance of Cu-$TaS_2$ interconnects at three different process-temperature corners. The main contributions of this work are summarized as follows.

- We model copper interconnects with $TaS_2$ barrier/liners based on temperature-dependent 3D simulations and present the implications of replacing standard barrier/liners (TaN/Ta) with ultra-thin $TaS_2$ on $R_{METAL}$ and $R_{VIA}$.
- We perform a comprehensive analysis of the impact of $TaS_2$-augmented interconnects on the circuit performance of AES at 7nm technology node vis-à-vis standard interconnects.
- We analyze the effect of vertical conductivity of $TaS_2$ on AES performance and provide guidelines for the further exploration of this promising interconnect technology.

## II. Background

### A. $TaS_2$-Augmented Interconnects

The schematic of a typical modern interconnect structure is illustrated in Fig. 1. Copper in both the metal lines and the vias is enclosed within layers comprising barrier (TaN) and liner (Ta) materials on all sides but the top. In the state-of-the-art interconnects, the barrier/liner thickness can be as thick as 2nm each (with 50% sidewall coverage for TaN barrier). With the shrinking footprint of the interconnects, the percentage of the copper layer (where the active conduction takes place) becomes significantly smaller. In the local level (M1-M3) line metals and vias, Cu conduction area for 7nm technology node is only 61%. The conduction area decreases further as the technology is scaled. This alarming trend continues to be worsened because of a significant increase in the resistivity (not just resistance) as the cross-sectional area decreases. This is primarily due to sidewall scattering. As a mitigation of this issue, $TaS_2$ provides a potential solution. The use of $TaS_2$ as the barrier and liner for advanced interconnects was proposed in [9]. This work demonstrates that $TaS_2$ can serve as an effective barrier (to prevent copper electromigration) and as a liner (as it displays excellent adhesion). Furthermore, due to its 2D nature, an ultra-thin (~0.7nm) $TaS_2$ layer can be fabricated, thus alleviating the issues associated with TaN/Ta based barrier/liner, which have fundamental limitations in their thickness scaling. To add to the appeal of $TaS_2$, it serves both as a barrier and liner, and thus, a single ultra-thin layer of $TaS_2$ can replace both TaN and Ta, thus increasing the percentage of copper area for M1-M3 to 90% at 7nm technology node, as shown in Fig. 1. The analysis in [11] shows that this increase in copper area reduces the unit resistance of line metals by up to 49% (compared to a standard interconnect). Scaling down further to a 5nm technology node, the reduction in unit resistance is as large as 58%.

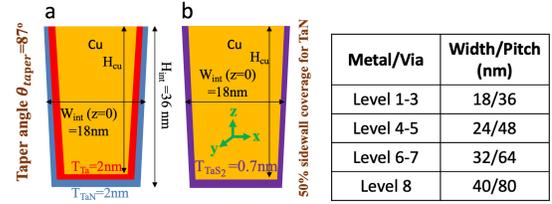

Fig. 1. a) Conventional metal/via  b) $TaS_2$-augmented metal/via. The table alongside shows the metal/via dimensions used in this study.

| Metal/Via | Width/Pitch (nm) |
|---|---|
| Level 1-3 | 18/36 |
| Level 4-5 | 24/48 |
| Level 6-7 | 32/64 |
| Level 8 | 40/80 |

However, such promising characteristics of $TaS_2$ come with their own issues. The vertical conductivity of $TaS_2$ is significantly small (despite its ultra-thin geometry) compared to the conventional TaN/Ta combination. The experimental data in [9], based on *ex-situ* measurements show a vertical resistivity of 20 $\Omega^*\mu m$. (Note, the horizontal resistivity of $TaS_2$ is measured to be 1.5 $\Omega^*\mu m$). The vertical resistivity may be even larger with $TaS_2$ integrated in the interconnect structures (which needs further experimental investigation). The small vertical conductivity is particularly critical for vias, in which the current flow is vertical (Fig. 2). Thus, while the increase in the copper percentage in Cu-$TaS_2$ interconnect enhances the conductivity, the detrimental effects on $R_{VIA}$ can be a serious concern.

To analyze the Cu-$TaS_2$ interconnects in a systematic fashion and understand the circuit implications of the opposing trends (of increasing copper area and reduced vertical conductivity), we model the $TaS_2$-augmented interconnects copper interconnects in this paper using a 3D simulation framework considering surface and grain-boundary scattering mechanisms and incorporating temperature-dependence of resistivity. Using this model, we analyze the circuit-level impact of $TaS_2$-augmented interconnects considering AES benchmark at different process-temperature corners.

### B. Benchmarking Flow

To evaluate the circuit implications of $TaS_2$ interconnects, we follow the standard place-and-route (PnR) flow shown in Fig. 3 and [11]. We use an Advanced Encryption Standard (AES) circuit [14] as a case study. The AES circuit is well-known for being interconnect dominated, making the circuit performance changes due to TaS$_2$-interconnects more obvious. The AES circuit RTL is synthesized using the commercial EDA tool Cadence Genus and ASAP7 7nm PDK [15]. In the case study presented in this work, we utilize the low threshold voltage (LVT) transistors targeting high circuit performance. In this PDK, there are Fast-Fast (FF), Typical-Typical (TT), and Slow-Slow (SS) corners corresponding to 0 °C, 25 °C and 100 °C, respectively. We analyze the AES benchmark considering these three process-temperature corners. The logic equivalence check is performed using Cadence Conformal LEC. The PnR is conducted using Cadence Innovus and the design rule check (DRC) and layout verification against schematic (LVS) are conducted using Cadence Virtuoso. The system verification is performed employing Questa Sim.

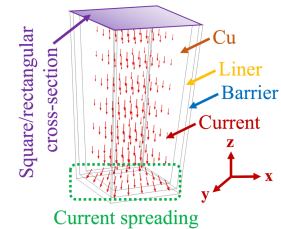

Fig. 2. Conventional Via struture in 3D, showing current flow vectors and current spearding effect.

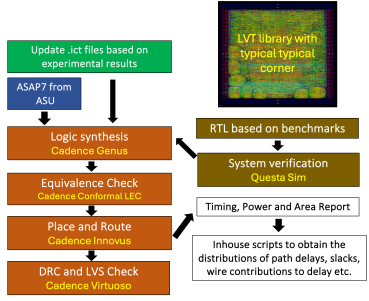

Fig. 3. Standard PnR flow used in this study, with fully routed AES circuit alongside.

### III. TEMPERATURE-DEPENDENT MODELING OF TaS$_2$-AUGMENTED INTERCONNECT TECHNOLOGIES

*A. Resistivity Model and 3D Simulation Framework*

The resistivity model employed in this study comprises two distinct components. First, surface scattering is represented by a "cosh" based model proposed in [16]. This empirical model incorporates space-dependent surface scattering, essential for accurately depicting modern interconnect topologies. Metal and via interconnects exhibit non-trivial current transport, primarily due to the current spreading effect (Fig. 2) associated with their tapered and multi-layered structures. Capturing the spatial dependency of conductivity is thus critical for understanding this complex current transport phenomenon. Second, the grain boundary scattering is modeled using the Mayadas-Shatzkes (MS) theory [13]. These two scattering mechanisms are combined through Matthiessen's rule to form the foundation of our resistivity model. We utilize approach to model both Cu-TaS$_2$ and standard interconnects (for comparison).

We further incorporate temperature dependence in the conductivity of Cu, Ta, TaN and TaS$_2$. We model the bulk resistivity ($\rho_0$) and electron mean free path ($\lambda_0$) as functions of the temperature (*T*). The physical relationship between $\rho_0$ and *T* is defined in [17] as follows:

$$\rho_0(T) = q_1 + \left(\frac{m_{eff}}{2}\right)^{\frac{1}{2}} \frac{q_2}{\Theta} \left(\frac{T}{\Theta}\right)^5 \int_0^{\Theta/T} \frac{z^5}{(e^z - 1)(1 - e^{-z})} dz \quad (1)$$

Here, $q_1$ is the residual resistivity, $m_{eff}$ represents the effective mass of the electron, $\Theta$ denotes the Debye temperature in Kelvin, and $q_2$ is a fitting parameter. For Cu, this model is calibrated against the experimental results presented in [18] (Fig. 4), yielding a residual resistivity of $q_1 = 0.02 \Omega$-nm and $q_2 = 43.05$ GΩ-m*K/kg$^{1/2}$. To account for temperature dependence of $\lambda_0$, we utilize the fact from electron gas theory that $\lambda_0 \rho_0 = 6.6 \times 10^{-16}$ for Cu and is independent of *T*. Consequently, we obtain:

$$\lambda_0(T) = \lambda_0(T = 25^0C)\rho_0(T = 25^0C)/\rho_0(T) \quad (2)$$

According to simplified Fuchs-Sondheimer (FS) theory [12], the contribution of surface scattering to resistivity is nearly temperature independent. Therefore, we do not include any temperature dependence in the empirical "cosh" based model from [16] which models surface scattering. For this model, we use the same experimentally calibrated parameters as in [16].

The grain boundary scattering in Cu, is temperature dependent, as it is influenced by $\lambda_0(T)$. To model this, we apply the Mayadas-Shatzkes (MS) theory [13], incorporating modifications to account for temperature dependence as shown in (3).

$$F_{MS}(T) = \left[1 - \frac{3\alpha(T)}{2} + 3\alpha(T)^2 - 3\alpha(T)^3 \ln\left(1 + \frac{1}{\alpha(T)}\right)\right]^{-1} \quad (3)$$

where $\alpha(T) = \lambda_0(T)R/[d_{grain}(1 - R)]$, $d_{grain}$ is the minimum grain size and *R* is the grain boundary reflection coefficient. We calibrate *R* (*R* = 0.135) to align the grain boundary scattering contribution with that reported in [16].

To obtain the resistivity-temperature relationship for TaS$_2$, TaN and Ta (with resistivity of 3 Ω*μm and 2 Ω*μm at room temperature respectively [11]), we employ a first-order linear approximation. This approach is considered appropriate for the temperature range utilized in this study, which spans 0°C to 100°C. The model parameters (such the thermal coefficient of resistance or TCR) for TaS$_2$, TaN and Ta are determined through calibration against experimental data sourced from [19], [20] and [21], respectively.

Building upon our 3D COMSOL interconnects modeling framework [11], we incorporate the temperature dependent resistivity model for various materials. This framework captures critical interconnect modeling aspects, including realistic tapered structure, space-dependent surface scattering, grain boundary scattering, and their temperature dependencies.

The width and pitch utilized in this work are illustrated in Fig. 1, sourced from ASAP7 PDK [15]. The interconnects feature an aspect ratio of 2 and a taper angle of 87° [22]. Conventional materials such as TaN and Ta exhibit a negative TCR, as reported in [20] and [21], indicating that their resistivity decreases with increasing temperature. Conversely, TaS$_2$ and copper display a positive TCRs [19], affecting both in-plane and out-of-plane resistivity. With these non-trivial temperature-dependencies of different materials, it becomes important to evaluate the overall trade-offs at the circuit level, as discussed in subsequent sections.

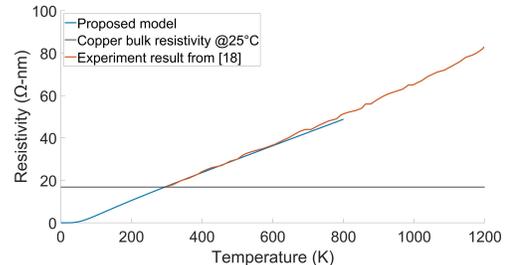

Fig. 4. Cu temperature dependent resistivity model calibration with [18], showing the model has a good fit with experimental results from 300K to 500K.

In this study, we conduct a performance comparison of TaS$_2$-augmented interconnects with two conventional interconnect structures ("conventional" and "scaled liner"). Recall, the first structure, referred to as "conventional", comprises a 2 nm layer of TaN followed by a 2 nm layer of Ta. The second structure, referred to as "scaled liner", involves a scaled-down liner where the TaN layer remains at 2 nm, but the Ta layer is reduced to 1nm. Liner scaling is a promising approach being explored to reduce the overall interconnect resistance. We build 3D temperature-dependent COMSOL models for all the three structures (conventional, scaled line and TaS$_2$-augmented) and incorporate them in the synthesis and place-and-route flow. In this work, we employ three different corners, corresponding to the Fast-Fast (FF) 0°C, Typical-Typical (TT) 25°C, and Slow-Slow (SS) 100°C, for the low $V_T$ (LVT) library from ASAP7 7nm PDK [15]. The temperature-dependent interconnect resistance values from our 3D COMSOL simulations are utilized in this flow to analyze the circuit performance.

IV. BENCHMARKING RESULTS AND CIRCUIT-LEVEL ANALYSIS

*A. Resistance Comparison at Different Temperatures*

The interconnect resistances obtained from our model at different temperatures are listed in Table I. The values for the conventional interconnects are similar to those reported in previous studies [10], [23] by other research groups.

For lower-level metal interconnects, TaS$_2$-augmented interconnects show line metal M1-M3 resistance reduction by 47%-49% compared to conventional and 31%-32% compared to scaled liner with the highest reduction observed at 0°C. For M8-M9 metal interconnects, 18%-21% reduction in resistance is observed. This substantial reduction in unit resistance is mainly attributed to the larger Cu conduction area achieved by utilizing a thin TaS$_2$ barrier/liner. For the lower-level vias, the TaS$_2$-augmented vias exhibit resistance reduction by 15% at 0°C but show a 1% increase in resistance at 100°C compared to conventional interconnects. However, at the higher level (see V8), the TaS$_2$-augmented vias exhibit up to 23% increase in resistance. This phenomenon can be attributed to the larger dimensions of high level vias compared to those at lower level. The increased via size reduces the contribution of Cu to the overall resistance (due to reduction in surface scattering), thereby enhancing the impact of the higher resistivity of TaS$_2$ on the total resistance.

The trends with respect to the temperature show an increase in line metal resistance with increasing temperature for all the interconnect structures. This is mainly due to positive TCR of Cu. (Note, the current from through the barrier/liner in line metals is negligible due to their much higher resistance compared to Cu). For TaS$_2$-augmented vias, the resistance increases with temperature due to the overall effect of the positive TCR for TaS$_2$ and Cu. On the other hand, for conventional and scaled liner vias, the temperature-dependence is mild which is due to the opposing effects of the positive TCR of Cu and the negative TCR of TaN and Ta. The largest benefits of TaS$_2$ are observed at 0°C as TCR for TaS$_2$ is positive while that of TaN and Ta (conventional liners/barriers) is negative.

TABLE I.
BEOL RESISTANCE WITH DIFFERENT TEMPERATURE

| Metal (Ω/μm) Via (Ω) | Conventional | | | Scaled Liner | | | TaS$_2$-augmented | | |
|---|---|---|---|---|---|---|---|---|---|
| Temp (°C) | 0 | 25 | 100 | 0 | 25 | 100 | 0 | 25 | 100 |
| M1-M3 | 236 | 243 | 253 | 177 | 182 | 192 | 121 | 125 | 133 |
| V1-V3 | 78 | 78 | 77 | 55 | 56 | 55 | 66 | 69 | 78 |
| M4-M5 | 82 | 85 | 91 | 66 | 68 | 74 | 50 | 52 | 57 |
| V4-V5 | 39 | 39 | 39 | 30 | 30 | 30 | 37 | 39 | 44 |
| M6-M7 | 30 | 31 | 35 | 26 | 27 | 30 | 20 | 22 | 25 |
| V6-V7 | 21 | 21 | 20 | 16 | 16 | 16 | 13 | 22 | 25 |
| M8-M9 | 14 | 15 | 17 | 13 | 14 | 16 | 11 | 12 | 14 |
| V8 | 13 | 13 | 13 | 10 | 10 | 10 | 13 | 14 | 16 |

To summarize, TaS$_2$-augmented interconnects significantly reduce line metal resistivity across all metal layers, with the largest reduction occurring at lower temperatures. Conversely, TaS$_2$-augmented lower-level vias exhibit a moderate to marginal advantage over conventional vias, with most benefits at lower temperature. High-level vias perform worse than conventional and scaled liner interconnects, especially at higher temperatures.

*B. Full Replacement of Conventional Interconnects with TaS$_2$-augmented interconnects.*

Following the benchmarking flow shown in Fig. 3, we compare the circuit performance for the TaS$_2$-augmented interconnects with the conventional and scaled liner interconnects under three temperature conditions. The methodology for the circuit analysis is outlined as follows. First, we execute the entire benchmarking flow for conventional, scaled liner, and TaS$_2$-augmented interconnects using SS corner at 100°C to obtain the routing for the worst-case process/temperature corner for timing. Subsequently, we replace the standard cell library at SS 100°C with the TT 25°C and interconnect resistances @ 100°C with those @ 25°C to determine the effective frequency at TT 25°C but using the same SS corner (worst-case design) routing (in alignment with the standard design process). The same procedure is applied for FF corner 0°C using the SS routing. Table II shows the effective frequency at different temperatures. At 100°C/SS corner, the TaS$_2$-augmented interconnects exhibit comparable performance (within 1% of effective frequency) with respect to the conventional interconnects. Even at 25°C/TT corner, comparable performance between two interconnects (within 0.6%) is observed. Compared to scaled liner, TaS$_2$-augmented interconnects exhibit some performance degradation at SS and TT corners. However, at 0°C/FF corner, TaS$_2$-augmented interconnects demonstrate the most improvement, with a 10.6% increase in effective frequency compared to the conventional interconnect and a 7 % improvement compared to the scaled liner interconnects.

From these trends, we make two observations. First, the effective frequency shows higher temperature sensitivity for TaS$_2$-augmented interconnects with an increase of 1.94 GHz (versus 1.6 GHz for conventional and 1.66 GHz for scaler liner) from 100°C to 0°C. Second, TaS$_2$-augmented interconnects exhibit better circuit level performance at lower temperatures compared to conventional and scaled liner. Recall, the conventional interconnects, which utilize TaN/Ta barriers/liner layers, exhibit a complex temperature-dependent effects due to the negative TCR of TaN/Ta and positive TCR of Cu. These opposing behavior results in a net reduction of temperature sensitivity in conventional/scaled liner interconnects. On the other hand, for the TaS$_2$-augmented interconnects, both TaS$_2$ and Cu exhibit positive TCRs. This alignment in TCR behavior implies that at lower temperature, both materials achieve their minimum resistivity. Consequently, TaS$_2$-augmented interconnects promise a better circuit-level performance at lower temperatures.

TABLE II.
CIRCUIT LEVEL RESULTS FOR DIFFERENT TEMPERATURE

|  | Effective Freq (GHz) @ 100°C (SS) | Effective Freq (GHz) @ 25°C (TT) | Effective Freq (GHz) @ 0°C (FF) |
|---|---|---|---|
| Conventional | 2.0 | 3.01 | 3.6 |
| Scaled Liner | 2.06 (3%) | 3.1 (3%) | 3.72 (3.3%) |
| TaS$_2$-augmented | 2.04 (1%) | 2.99 (-0.6%) | 3.98 (10.6%) |
| TaS$_2$-augmented 1-3 Conventional 4+ | 2.06 (3%) | 3.05 (1.3%) | 4.05 (12.5%) |
| TaS$_2$-augmented 1-5 Conventional 6+ | 2.07 (3.5%) | 3.05 (1.3%) | 4.03 (12%) |
| TaS$_2$-augmented 1-7 Conventional 8+ | 1.98 (1%) | 3.0 (0.3%) | 3.65 (1.4%) |

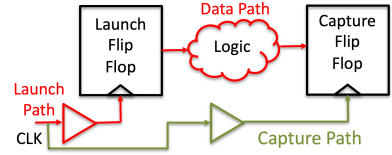

Fig. 5. Simplified schematic showing the data and capture paths

*C. Hybrid Routing Scheme*

From Table I, we observe that higher level TaS$_2$-augmented vias exhibit worse performance compared to their conventional counterparts due to high TaS$_2$ vertical resistivity, especially for higher level vias. To address this disadvantage, we analyze a hybrid routing scheme by partially replacing conventional with TaS$_2$-augmented interconnects (Table II).

Using the SS process corner standard cells (worst case scenario), when we replace only level 1-3 layers of interconnects with TaS$_2$-augmented interconnects, 2.06 GHz effective frequency is achieved, which is a 3% improvement compared to the conventional interconnects. This corresponds to a 1% higher performance improvement compared to the circuit fully routed with TaS$_2$-augmented interconnects. Furthermore, compared to conventional design, replacing level 1-5 conventional interconnects with TaS$_2$-augmented interconnects results in a 3.5% performance improvement, achieving a frequency of 2.07 GHz. Conversely, extending the replacement to levels 1-7 resulted in a 1% performance reduction (with frequency of 1.98 GHz). For 25 °C/TT corner, we observe similar results as 100 °C/SS corner.

For the 0 °C/FF corner, when we replace only level 1-3 layers of interconnects with TaS$_2$-augmented interconnects, 4.05 GHz effective frequency can be achieved, which is a 12.5% improvement compared to the conventional interconnects. Further replacement of higher level (level 1-5 and 1-7) interconnects with TaS$_2$-augmented interconnects yields diminishing improvements (12% and 1.4%, respectively) compared to conventional interconnects. These findings suggest that a partial replacement strategy, specifically targeting lower levels of the interconnect hierarchy, offers a promising approach to enhancing circuit level performance with TaS$_2$-augmented interconnects.

*D. Analysis of the Timing Paths*

To gain further insights into these trends, we perform an analysis of the path delays at different temperatures with a focus on conventional and TaS$_2$-augmented interconnects. For the latter, we consider the full replacement of all metal layers. For our analysis, let us consider a simplified schematic for a stage of a pipeline as shown in Fig. 5, which illustrates the launch path delay (CLK to launch flip-flop), data path delay (propagation through logic gates), and capture path delay (CLK to capture flip-flop). The operational frequency of the circuit is inversely related to the launch path delay and data path delay.

Conversely, a higher capture path delay can mitigate setup time constraints by increasing the clock skew and the frequency. The critical path, defined as the path with the highest delay, dictates the clock frequency. To assess the impact of TaS$_2$-augmented interconnects, we examine the delay for different paths, with a particular focus on the critical path. Key parameters for each path include data required time (associated with capture path delay minus setup time), data arrival time (a function of launch path delay plus data path delay), and worst negative slack (WNS - defined as the difference between data required time and data arrival time). Using our in-house scripts, we extract timing information for a fully routed AES circuit, comprising over 5000 paths. For the rest of the study, we combine the launch path and data path and refer them together as data path.

To analyze the performance between conventional and TaS$_2$-augmented interconnects, we present the results for FF corner @ 0°C and SS corner @ 100°C. (Recall, the routing is based on the worst case design i.e. SS@100°C). We analyze the top 100 critical paths based on WNS. To understand the implications of using TaS$_2$-augmented interconnects, let us analyze the capture path delays (which dictates the clock skew and the data required time) and the data path delays (which determine the data arrival time). Fig. 6a and 6b show that the data required time for TaS$_2$-augmented interconnects is *moderately* less than that for conventional interconnects for both the FF and SS corners. Conversely, TaS$_2$-augmented interconnects exhibit a *considerable* reduction in data arrival time for both the FF and SS corners, as show in Fig. 7a and 7b. The reduction in the data required time and data arrival time is due to lower interconnect resistance. However, to understand why the data arrival time is more sensitive to interconnect resistance compared to the data required time, we present the following possible explanation. Recall that the data required time is associated with the capture path delay, while the data arrival time is associated with the data path delay (i.e. data path combined with the launch path). The capture path is usually designed by inserting buffers appropriately so that the clock slew rate is higher than a critical value. Higher resistance of conventional interconnects (compared to TaS$_2$-augmented interconnects) does result in increase in the capture path delay; however, buffer insertion (optimized for each type of interconnect individually) partially (but considerably) offsets the effect of the interconnect resistance on delay. Even for the data path, optimizations are performed with the logic gates and the routing. However, these optimizations are based on less constrained slew rate targets for the signals (compared to the clock in the capture paths) and therefore, have a lower impact of offsetting the impact of interconnect resistance on delay. Consequently, we observe a much larger delay reduction due to TaS$_2$-augmented interconnects in the datapath, as compared to the capture path.

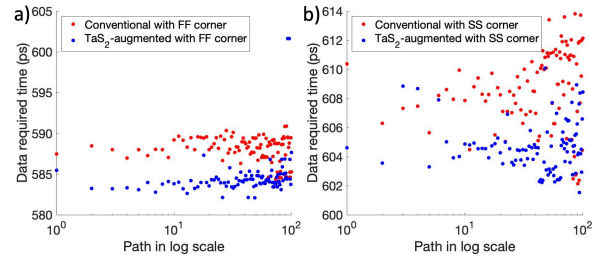

Fig. 6. a) Data required time with FF corner b) data required time with SS corner time comparison using worst case routing @100 °C.

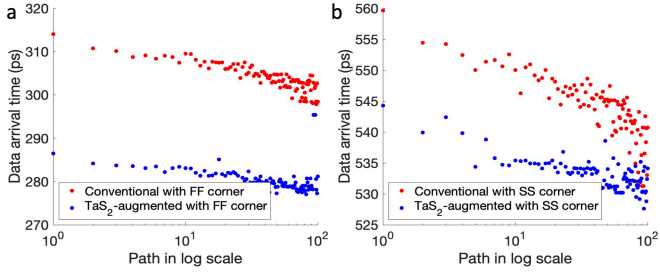

Fig. 7. a) Data arrival time with FF corner b) data arrival time with SS corner time comparison using worst case routing @100 °C.

Quantifying the impact of wire resistance on the path delays, we observe <5.4% reduction in data required time (on an average) for TaS$_2$-augmented interconnects compared to conventional interconnects at FF corner $0^0$C (Fig. 6a). Note that this negatively impacts the system performance, as the clock skew is reduced thereby tightening the setup time constraints. The SS $100^0$C corner results are shown in Fig. 6b, showing reduction in data required time by 1% for TaS$_2$-augmented interconnects. In this corner, higher effective resistance of the logic gates (compared to the FF corner) reduces the sensitivity of path delays to the wire resistance.

Now, if we analyze the data arrival times in Fig. 7, we observe that TaS$_2$-augmented interconnects achieve up to 7.7% reduction (on an average) due to decreased launch and data path delay delays relative to conventional interconnects (Fig. 7a). Smaller (2%) reduction is observed for the SS corner (Fig. 7b).

If we focus on the topmost critical path (with the lowest WNS and which dictates the effective frequency of the system), we observe that TaS$_2$-augmented interconnects in the FF corner 0°C reduce the data required time and data arrival time by 3.4% and 8.8% (1% and 3% for SS corner), respectively. Hence, the dual impact of TaS$_2$-augmented interconnects is to reduce the data arrival time but also the data required time. While the former contributes in enhancing the system performance, the latter partially offset the benefits of TaS$_2$-augmented interconnects. However, since the impact of reduced resistance of TaS$_2$-augmented interconnects is higher on the data arrival time (compared to the data required time – as discussed above), an overall increase in the effective frequency is achieved by using the TaS$_2$-augmented interconnects, especially in the FF $0^0$C corner (as quantified in the previous sub-section).

*E. Wire Delay Analysis*

To understand the implications of TaS$_2$-augmented interconnects further, we analyze the wire delay contribution in the data and capture paths. For this, we extract wire delays for different paths using our in-house scripts. Fig. 8 shows the distribution of % wire delay (wire delay/total delay × 100%) in the data paths and capture paths corresponding the top 100 critical paths of the AES circuit. Notably, the wire delay contribution for TaS$_2$-augmented interconnects shift towards lower values for both the data and capture paths. This shift is attributed to the lower resistance of TaS$_2$-augmented interconnects, which significantly reduces wire delay.

The analysis of the top 100 paths in both designs (for FF @0°C) shows that TaS$_2$-augmented interconnects result in an average wire delay contribution of 11% for the data path, compared to 17% for the conventional (i.e. 0.65X reduction in % wire delay in TaS$_2$-augmented interconnects with respect to the conventional). For the capture path, TaS$_2$-augmented interconnects show an average wire delay contribution of 28% compared to 33% for the conventional (i.e. 0.84X reduction in % wire delay contribution). These results underscore the potential of TaS$_2$-augmented interconnects to enhance overall circuit performance by mitigating wire delay. Also, note that the impact on the lower interconnect resistance on the wire delay contribution is more for the data path than the capture path, which is in alignment with our previous discussion on data required time and data arrival time. As discussed before, the reduction in capture path delay offsets some of the benefits at circuit level, which must be taken into account while understanding the implications of low-resistance interconnects on the circuit performance.

## V. THE IMPACT OF VERTICAL RESISTIVITY OF TaS$_2$ INTERCONNECT RESISTANCE AND CIRCUIT PERFORMANCE

As noted earlier, one of the shortcomings of TaS$_2$ material is significantly higher vertical resistivity compared to conventional materials. This has substantial impact on the via resistance. Our analysis, so far, is based on the value (baseline 1x) from the experimental ex-situ results from [9]. In this section, we perform a what-if analysis to understand the effect of varying TaS$_2$ vertical resistivity.

Following the approach in Section III, we modify the vertical resistivity of TaS$_2$ by factors of 0.5x, 2x and 4x relative to the baseline value at $0^0$C, $25^0$C and $100^0$C. The resulting BEOL parameters are presented in Table III. Next, we determine the impact of the vertical resistivity of TaS$_2$ on the circuit performance by using the parameters in Table III in our synthesis and place-and-route flow. Note the routing is performed for SS corner 100°C (worst-case design) *individually* for each vertical resistivity and the effective frequency is obtained. Here, we limit our discussions of effective frequency to SS corner only. Our goal is to identify the break-even point for the vertical resistivity of TaS$_2$ below which TaS$_2$-augmented interconnects offer circuit performance improvements over conventional interconnects.

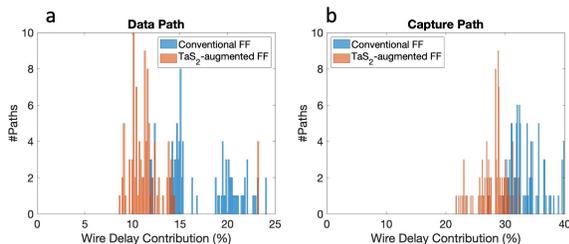

Fig. 8. a) Data b) capture path wire delay histogram using FF corner @ 0 °C.

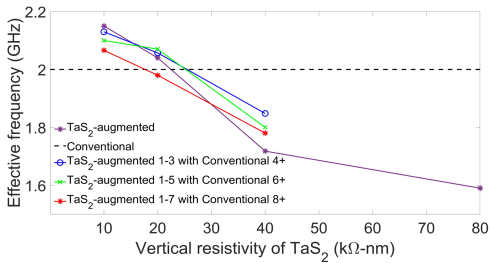

Fig. 9. Effective frequency vs TaS$_2$ vertical resistivity @100 °C for full replacement and hybrid routing scheme.

The results are shown in Fig. 9. As the vertical resistivity of the TaS$_2$ increases, the effective frequency decreases steadily. To achieve performance benefits from TaS$_2$, its vertical resistivity must be lower than 22 kΩ-nm.

Additionally, we implemented the hybrid routing scheme for different vertical resistivities of TaS$_2$. It can be observed that the hybrid routing mildly relaxes the requirement for TaS$_2$ vertical resistivity to achieve performance benefits, with the break-even point increasing to 25 kΩ*nm when TaS$_2$-augmented interconnects replace conventional interconnect at levels 1-3 and levels 1-5. Further replacement of higher level interconnects with TaS$_2$-augmented interconnects, shown by the red line, results in a more constrained break-even point. These results underscore our conclusions that replacing the lower-level metal layers with TaS$_2$-augmented interconnects is the most reasonable approach.

TABLE III.
VIA RESISTANCE WHEN VARY VERTICAL RESISTIVITY AND TEMPERATURE

| Via (Ω) | 0.5x | | | 1x | | | 2x | | | 4x | | |
|---|---|---|---|---|---|---|---|---|---|---|---|---|
| | 0 | 25 | 100 | 0 | 25 | 100 | 0 | 25 | 100 | 0 | 25 | 100 |
| V1-V3 | 40.6 | 42 | 47 | 66 | 69 | 78 | 118 | 124 | 140 | 220 | 231 | 264 |
| V4-V5 | 22.7 | 23.6 | 26 | 37 | 39 | 44 | 66 | 69 | 79 | 124 | 130 | 149 |
| V6-V7 | 13 | 13 | 15 | 13 | 22 | 25 | 37 | 39 | 45 | 70 | 73 | 84 |
| V8 | 8.1 | 8.5 | 9.6 | 13 | 14 | 16 | 23 | 25 | 29 | 45 | 47 | 54 |

## IV. CONCLUSION

In this work, we evaluated the impact of TaS$_2$-augmented interconnects on circuit level performance with temperature dependency taken into account. Utilizing the interconnect models capturing surface scattering and grain-boundary scattering and a circuit benchmarking flow, we conducted a full place-and-route (PnR) of the AES circuit across three distinct process and temperature corners defined by the ASAP7 7nm PDK [15]. Our modeling results show that TaS$_2$-augmented line metal interconnects exhibit a substantial resistance reduction of 31%-49% compared to both conventional and scaled liner line metal interconnects, with the most significant reduction observed at 0°C. On the other hand, TaS$_2$-augmented interconnect vias experience a moderate resistance decrease of 15% compared to conventional vias at 0°C and 25°C, but an increase of 1% at 100°C. When compared to scaled liners, TaS$_2$-augmented vias show higher resistance across the entire temperature range. Our circuit level benchmarking results reveal that compared to the conventional interconnects, TaS$_2$-augmented interconnects achieve comparable performance at the SS 100$^0$C corner (worst-case scenario), but enhance the effective frequency by a remarkable 10.6% at the FF 0$^0$C corner. Further, for the SS corner, the vertical resistivity of the TaS$_2$ material must remain below 22 kΩ*nm to realize performance benefits over conventional interconnects. Additionally, we observe that a partial replacement of conventional interconnects with TaS$_2$-augmented interconnects yields higher performance boost compared to full replacement. The most significant improvement is achieved by replacing local-level interconnects (level 1-3) with TaS$_2$-augmented interconnects. In this scenario, the vertical resistivity break-even point increases to 25 kΩ*nm.

REFERENCES

[1] M. Horowitz, E. Alon, D. Patil, S. Naffziger, R. Kumar, and K. Bernstein, "Scaling, power, and the future of CMOS," in *IEEE InternationalElectron Devices Meeting, 2005. IEDM Technical Digest.*, 2005, pp. 7 pp. – 15. doi:10.1109/IEDM.2005.1609253.
[2] R. Brain, "Interconnect scaling: Challenges and opportunities," in *2016 IEEE International Electron Devices Meeting (IEDM)*, 2016, pp. 9.3.1-9.3.4. doi: 10.1109/IEDM.2016.7838381.
[3] A. P. Jacob, R. Xie, M. G. Sung, L. Liebmann, R. T. P. Lee, and B. Taylor, "Scaling Challenges for Advanced CMOS Devices," *International Journal of High Speed Electronics and Systems* , p. 1740001, 2017, [Online]. Available: www.worldscientific.com
[4] J. S. Chawla, F. Gstrein, K. P. O'Brien, J. S. Clarke, and D. Gall, "Electron scattering at surfaces and grain boundaries in Cu thin films and wires," *Phys Rev B Condens Matter Mater Phys*, vol. 84, no. 23, Dec. 2011, doi: 10.1103/PhysRevB.84.235423.
[5] W. L. Wang *et al.*, "The Reliability Improvement of Cu Interconnection by the Control of Crystallized α-Ta/TaNx Diffusion Barrier," *J Nanomater*, vol. 2015, 2015, doi: 10.1155/2015/917935.
[6] D. Edelstein *et al.*, "A high performance liner for copper damascene interconnects," in *Proceedings of the IEEE 2001 International Interconnect Technology Conference (Cat. No.01EX461)*, 2001, pp. 9–11. doi: 10.1109/IITC.2001.930001.
[7] N. Bekiaris *et al.*, "Cobalt fill for advanced interconnects," in *2017 IEEE International Interconnect Technology Conference (IITC)*, 2017, pp. 1–3. doi: 10.1109/IITC-AMC.2017.7968981.
[8] D. Wan *et al.*, "Subtractive Etch of Ruthenium for Sub-5nm Interconnect," in *2018 IEEE International Interconnect Technology Conference (IITC)*, 2018, pp. 10–12. doi: 10.1109/IITC.2018.8454841.
[9] C. L. Lo *et al.*, "Opportunities and challenges of 2D materials in back-end-of-line interconnect scaling," Aug. 28, 2020, *American Institute of Physics Inc.* doi: 10.1063/5.0013737.
[10] D. E. Shim, V. Huang, X. Chen, S. K. Gupta, and A. Naeemi, "A Comprehensive Modeling Platform for Interconnect Technologies," *IEEE Trans Electron Devices*, vol. 70, no. 5, pp. 2594–2599, 2023, doi: 10.1109/TED.2023.3261828.
[11] X. Chen, C.-L. Lo, M. C. Johnson, Z. Chen, and S. K. Gupta, "Modeling and Circuit Analysis of Interconnects with TaS2 Barrier/Liner," in *2021 Device Research Conference (DRC)*, 2021, pp. 1–2. doi: 10.1109/DRC52342.2021.9467160.
[12] E. H. SONDHEIMER, "Influence of a Magnetic Field on the Conductivity of Thin Metallic Films," *Nature*, vol. 164, no. 4178, pp. 920–921, 1949, doi: 10.1038/164920a0.
[13] A. F. Mayadas and M. Shatzkes, "Electrical-Resistivity Model for Polycrystalline Films: the Case of Arbitrary Reflection at External Surfaces," *Phys Rev B*, vol. 1, no. 4, pp. 1382–1389, Feb. 1970, doi: 10.1103/PhysRevB.1.1382.
[14] Joachim Strömbergson, "aes." Accessed: Nov. 30, 2023. [Online]. Available: https://www.github.com/secworks/aes
[15] L. T. Clark *et al.*, "ASAP7: A 7-nm finFET predictive process design kit," *Microelectronics J*, vol. 53, pp. 105–115, Jul. 2016, doi: 10.1016/j.mejo.2016.04.006.
[16] I. Ciofi *et al.*, "Impact of Wire Geometry on Interconnect RC and Circuit Delay," *IEEE Trans Electron Devices*, vol. 63, no. 6, pp. 2488–2496, Jun. 2016, doi: 10.1109/TED.2016.2554561.
[17] E. H. Sondheimer, "The mean free path of electrons in metals," *Adv Phys*, vol. 1, no. 1, pp. 1–42, 1952, doi: 10.1080/00018735200101151.
[18] L. Abadlia, F. Gasser, K. Khalouk, M. Mayoufi, and J. G. Gasser, "New experimental methodology, setup and LabView program for accurate absolute thermoelectric power and electrical resistivity measurements between 25 and 1600 K: Application to pure copper, platinum, tungsten, and nickel at very high temperatures," *Review of Scientific Instruments*, vol. 85, no. 9, Sep. 2014, doi: 10.1063/1.4896046.
[19] J. P. Tidman, O. Singh, A. E. Curzon, and R. F. Frindt, "The phase transition in 2H-TaS2 at 75 K," *Philosophical Magazine*, vol. 30, no. 5, pp. 1191–1194, 1974, doi: 10.1080/14786437408207274.
[20] A. Malmros, K. Andersson, and N. Rorsman, "Combined TiN- and TaN temperature compensated thin film resistors," *Thin Solid Films*, vol. 520, no. 6, pp. 2162–2165, Jan. 2012, doi: 10.1016/j.tsf.2011.09.050.
[21] M. Zhang, Y. F. Zhang, P. D. Rack, M. K. Miller, and T. G. Nieh, "Nanocrystalline tetragonal tantalum thin films," *Scr Mater*, vol. 57, no. 11, pp. 1032–1035, Dec. 2007, doi: 10.1016/j.scriptamat.2007.07.041.
[22] T. Lu and A. Srivastava, "Detailed electrical and reliability study of tapered TSVs," in *2013 IEEE International 3D Systems Integration Conference (3DIC)*, 2013, pp. 1–7. doi: 10.1109/3DIC.2013.6702350.
[23] V. Huang, D. E. Shim, J. Kim, S. Pentapati, S. K. Lim, and A. Naeemi, "Modeling and Benchmarking Back End Of The Line Technologies on Circuit Designs at Advanced Nodes," in *2020 IEEE International Interconnect Technology Conference (IITC)*, 2020, pp. 37–39. doi: 10.1109/IITC47697.2020.9515629.